\documentstyle[12pt]{article}

        \def\be{\begin{equation}}
        \def\ee{\end{equation}}
        
        \def\m{\mu}
        \def\s{\sigma}

\begin{document}

\title {\normalsize \bf " A Solvable Hamiltonian System " \\ Integrability and Action-Angle
Variables}

\author {V. Karimipour}
\begin{center}
\maketitle{\it Institute for Studies in Theoretical Physics and Mathematics\\
P. O. Box 19395-5531, Tehran , Iran \\ Department of Physics , Sharif
Uinversity of Technology \\P O Box 11365-9161, Tehran , Iran \\
} \end{center}  \vspace {12 mm}
\begin{abstract}
\ \ \ \ We prove\  that\ \  the \  dynamical \ \ system\ \  charaterized\  \ by \ the \ \ \ Hamiltonian
$ H = \lambda N  \sum_{j}^{N} p_j + \mu \sum_{j,k}^{N} {{(p_j p_k)}^{1\over 2}} \{ cos [ \nu ( q_j - q_k)] \} $
proposed and studied by Calogero [1,2] is equivalent to a system of {\it non-interacting} harmonic oscillators.
We find the explicit form of the conserved currents which are in involution. We
also find the action-angle variables and solve the initial value problem in simple form.
\end{abstract}

\section{Introduction}
Motivated by the discovery [3] that the dynamical system characterized by
the Hamiltonian
\be H = \sum_{j,k}^{N}{{(p_j p_k)}} exp \{ - \eta ( q_j - q_k) \} \ee
is completely integrable, Calogero [1] proposed a new Hamiltonian
system with the standard Poisson bracket and the Hamiltonian
\be H = \lambda N  P + \mu \sum_{j,k = 0}^{N-1} {{(p_j p_k)}^{1\over 2}} \{ cos [ \nu ( q_j - q_k)] \} \ee
\be P =  \sum_{j = 0}^{N-1} p_j  \ee
The equations of motions of the dynamical variables are
\be {d \over dt} q_j = \lambda N + \mu {p_j}^{{-1\over 2 }}\sum_{k=0}^{N-1}{p_k}^{{1\over 2 }}  \cos [ \nu(q_j - q_k )] \ee
\be {d \over dt} p_j = 2\mu \nu {p_j}^{{1\over 2 }}\sum_{k=0}^{N-1}{p_k}^{{1\over 2 }}  \sin [ \nu(q_j - q_k )] \ee
Then by way of a lenghty anlalysis, he succeeded to show that this Hamiltonian system is solvable, in the sense that
the above evolution equations can be solved in closed form. The final form of
his solution  take the following form:
$$ q_j (t) = q_j (0) + \lambda N t +  $$ \be \\ {\nu}^{-1} \arctan \Bigg
( { \sin \big ( \nu [ \mu N t + \alpha - q_j(0)] \big ) -
\sin \big ( \nu [  \alpha - q_j(0)] \big )\over
{ \cos \big ( \nu [ \mu N t + \alpha - q_j(0)] \big ) -
\cos \big ( \nu [ \alpha - q_j(0)] \big )}}  +
{ N\over A } [ p_j(0)]^{{1\over 2 }} \Bigg ) \ee

$$ p_j (t) = p_j (0) + 2 { A\over N }[ p_j(0)]^{{1\over 2 }}
\Bigg ( \cos \big ( \nu [ \mu N t + \alpha - q_j(0)] \big ) -
\cos \big (\nu [ \alpha - q_j(0)] \big )\Bigg ) + $$ \be
2 ({ A\over N })^2 [ 1 - \cos ( \nu \mu N t ) ] \ee

Where the constants $ A $  and $ \alpha $ are determined by the initial data as follows:
\be C(0) \equiv A \cos \alpha = \sum_{k=0}^{N-1}{{p_k}(0)}^{{1\over 2 }}  \cos [ \nu q_k(0) )] \ee
\be S(0) \equiv A \sin \alpha = \sum_{k=0}^{N-1}{{p_k}(0)}^{{1\over 2 }}  \sin [ \nu( q_k(0) )] \ee
This part of the analysis was done mainly by solving the equation of motion
for the functions $ C(t) $ and $ S(t) $  plus lenghty and at times cumbersome use of trigonometry. \\
He then went on to study the
" Behavior Near Equilibria " ( only in the case $ \lambda = 0 $ ) and by analysing
this system along the lines of the standard theory of small oscillations
, concluded that [1]:\\
" .... In the case $ \lambda = 0 $  considered here, the general motion is characterized
by a completely periodical behavior..."\  and \ ".... In view of the remarkable
simplicity of the motion characterizing the model, it is natural to conjecture
that there also exist a quantal version of this model and perhaps of some of its
generalizations which is solvable and features a spectrum whose discrete part is
equispaced....". The quantum version of this model was then discussed in ref.[2] .

This part of the analysis was done by heavy use of properties of some special
matrices introduced for this purpose in [1] .

The discovery of new integrable systems have always created excitement and
great activity in mathematical physics. These models have always proved to
have a hidden algebraic structure which is responsible for their integrability [4].
The study of this algebraic structure also helps us to generalize the original model
in many different ways. Motivated by the desire to understand the model (2) in
this framework we have studied this system from a purely algebraic point of view.

Our result is that when written in terms of suitable coordinates the Hamiltonian
(2) takes a very simple form, which is nothing but the Hamiltonian of N Non-Interacting harmonic
oscillators. All the results of Calogero [1] and Calogero and van Diejen [2] can then be
derived without doing any calculations. Furthermore we will find the integrals
of motion which are in involution, hence we prove the integrability of the
system and find the explicit form of the action-angle variables .

\section{ The New Dynamical Variables }

Define the variables
\be a_j = ({ p_j\over \nu} ) ^ {{1\over 2 }} e^{ i\nu q_j }\ee
\be a_j^{*} = ( {p_j\over \nu} ) ^ {{1\over 2 }} e^{ -i\nu q_j }\ee
Since the parameter $ \nu $ has the dimension of inverse lenght, the above variables will have the dimension
of (Action)$^{1\over 2}$ and will
have the following poisson brackets:
\be \{ a_j , a_k \} = \{ a_j^{*} , a_k^{*} \} = 0 \ \ \ \ \ \ \{ a_j , a_k^{*} \} = i \delta_{j,k} \ee
The Hamiltonian will then take the following form:
\be H = N\lambda \nu \sum_{j=0}^{N-1} a_j^{*} a_j  + \mu \nu  \sum_{j,k=0}^{N-1} a_j^{*} a_k  \ee
which can be put into the compact matrix form:
\be H = N \lambda \nu A^{\dagger} A + \mu \nu A^{\dagger} C A \ee
where \be A = \left( \begin{array}{l} a_1 \\ a_2 \\ . \\ . \\ a_{N-1}
\\ \end{array} \right)
 \ \ \ \ \ \ \ \ \ C = \left( \begin{array}{llllll} 1 & 1 & . & . &.& 1  \\
 1 & 1 & . & . &.& 1 \\ 1 & 1 & . & . &.& 1\\ . & . &. & . & . &.\\
1 & 1 & . & . & .& 1
\\ \end{array} \right)  \ee
where dot stands for 1. Having all of its elements equal to 1, the matrix  $ C $ has the following orthonormal eigenvectors:
\be C {\bf u}_{\alpha} = \xi_{\alpha} {\bf u}_{\alpha} \ \ \ \ \  \alpha = 0 , 1 , 2 , . . . .  N-1 \ee
where \be \xi_0 = N \ \ \  \xi_1 = \xi_2 = \xi_3 = . . . . \xi_{N-1} = 0  \ee and \be
({\bf u}_{\alpha})_s = {1\over {\sqrt N}}\ \ \omega ^ { \alpha s } \ \ \ \  with \ \ \ \ \  \omega = e^{{2\pi i \over N }} \ee
i.e:
\be {\bf u}_0  = \left( \begin{array}{l} 1 \\ 1 \\ 1 \\ . \\ . \\ 1 \\ \end{array} \right)
\ \ \ \ \ \ {\bf u }_1 = \left( \begin{array}{l} 1 \\ \omega \\ \omega^2 \\ . \\ . \\ \omega^N \\ \end{array} \right)
\ \ \ \ \ \ {\bf u }_2 = \left( \begin{array}{l} 1 \\ \omega^2 \\ \omega^4 \\ . \\ . \\ \omega^{2N} \\ \end{array} \right)
\ \ \ \ \ \ etc. \ee

Clearly the matrix R which diagonalizes the hermitian matrix C is unitary, with
\be R C R^{\dagger} \equiv D = diagonal ( N , 0 , 0 , 0 . . . 0 ) \ee
where the explicit form of $ R $ is given by $$ R_{\alpha, \beta } = ({\bf u}_{\alpha})_{\beta}= {1\over {\sqrt N}}\omega^{\alpha s } $$

Defining the new variables $ B = R A $ , i.e: \be b_{\alpha} = {1\over {\sqrt N}}\ \ \omega ^ { \alpha s } a_s \ \ \ \
b_{\alpha}^{*} = {1\over {\sqrt N}}\ \ \omega ^ { - \alpha s }a_s^{*} \ee we find:
\be H = N \lambda \nu B^{\dagger}B + \mu \nu B^{\dagger} D B \ee
or
\be H = N \lambda \nu \bigg ( b_0^{*} b_0 + b_1^{*} b_1 +  b_2^{*} b_2  + . . . . +  b_{N-1}^{*} b_{N-1} \bigg ) + \mu N \nu b_0^{*} b_0   \ee
It is essential that the new variables satisfy canonical commutation relations
which is a simple consequence of unitarity of R.
\be \{ b_{\alpha} , b_{\beta} \} = \{ b_{\alpha}^{*}
, b_{\beta}^{*} \} =  0 \ \ \ \ \ \ \ \ \{ b_{\alpha} ,  b_{\beta}^{*} \} = i \delta _{\alpha , \beta } \ee

Defining  now the quantities  $ I_{\alpha} \equiv b_{\alpha}^{*} b_{\alpha} $ which are global functions of the old coordiantes
we find :
\be \{ I_{\alpha} , I_{\beta }\} = 0  \ \ \ \ \ \ \forall \alpha , \beta \ee
The quantities $ I_{\alpha} $ are the N integrals of the motion, in involution with each other
and the Hamiltonian is a function of these integrals,i.e:
\be H = N \lambda \nu \bigg( I_0 + I_1 + I_2 + . . . . I_{N-1} \bigg) + \mu \nu N I_0 \ee
Therefore the system (2) is integrable in the Lioville sense.
The explicit form of the integrals $ I_{\alpha} $ is found from (10-11) and (21) to be :
\be I_{\alpha} = {1\over { 2 N \nu }} \sum_{k,k'=0}^{N-1}
{(p_k p_{k'})}^{1\over 2}\cos [ \nu ( q_{k'} - q_{k})+ { {2\pi \alpha}\over N} ( k' - k ) ] \ee

 These  functions ,
having the dimension of action are in fact the action variables. We can also
find the angle variables $ Q_{\alpha} $. These are:
\be Q_{\alpha} = {1\over 2i}  ln { b_{\alpha} \over b_{\alpha}^{*}} \ee
It is straightforward to check from (24) that the action-angle variables have canonical
poisson brackets:
\be \{ I_{\alpha } , I_{\beta}\} = \{ Q_{\alpha } , Q_{\beta}\} = 0 \ \ \ \ \ \ \ \ \ \{ Q_{\alpha } , I_{\beta}\} = \delta_{\alpha , \beta } \ee
The total momentum $ P $  has a simple expression in terms of the action variables:
\be P = \nu \bigg ( I_0 + I_1 + I _2 + . . . .  I_{N-1} \bigg ) \ee
It is now clear that the initial value problem can be solved in very simple and
closed form. The equations of motion are:
\be {d \over dt } b_{\alpha} = i \nu N \lambda b_{\alpha} \ \ \ \ \ \ \ \  \alpha\ne 0  \ee
\be {d \over dt } b_{0} = i \nu N ( \lambda+ \mu ) b_{0} \ \ \ \ \ \ee
with solutions
\be b_{\alpha}(t)  = b_{\alpha}(0)
e^{ i N \lambda t } \ \ \ \  \alpha \ne 0 \ee and
\be b_{0} (t) = b_{0}(0)
e^{i N ( \lambda+ \mu )t } \ \ \ \ \ \ee
Using the inverse transformation of (21) one obtains the time evolution of the
varibles $ (a_i,a_i^{*}) $ and hence of the original coordiantes and momenta. It is only for the difference between evolution of $ b_{\alpha \ne 0 } $ and $ b_{0} $ that
the varibles $ ( a_i,a_i^{*}) $ and hence $ q_i $ and $p_i $ have a complicated-looking evolution. Otherwise
( in the case $ \mu = 0 $ ) one will find that the variables $ ( a_i, a_i^{*}) $  have exactly the
same simple time evolution as the variable $ (b_i,b_i^{*}) $.

\section{Generalizations}

One can now generalize the system (2) in many different ways.
In fact from our analysis it is seen that if the matrix C in (14) is replaced
by any other hermitian matrix , none of the main results of this paper will change, except
the explicit form of the diagonalizing matrix. A good choice of the matrix
C which is perhaps much better than (15) on physical grounds , is the following:
\be C = \left( \begin{array}{lllll} 0 & 1 & . & . & {\s}  \\
1 & 0 & 1 & . & . \\ . & 1 & 0 & 1 & .\\ . & . & 1 & 0 & 1 \\
{\s} &. & . & 1 & 0
\\ \end{array} \right)  \ee  where dot stands for 0.
This matrix leads to the following Hamiltonian
$$ H = N\lambda \nu \sum_{j=0}^{N-1} a_j^{*} a_j  + \mu\nu \sum_{j=0}^{N-1} a_j^{*} a_{j+1}
+ a_{j+1}^{*} a_{j} + $$ \be \mu \nu \s ( a_1^{*} a_N + a_N^{*} a_1 ) \ee or
$$ H = N {\lambda }  \sum_{j}^{N} p_j + {\mu }\sum_{j = 0}^{N-1} {{(p_j p_{j+1})}^{1\over 2}} cos [ \nu ( q_j - q_{j+1})]  + $$
\be  \m \s (p_1 p_N)^{1\over 2} \cos [\nu (q_1 - q_N)] \ee which represents a system
with local ( nearest neighbor ) interaction , rather the long range interactions
implied by (2). Physically, systems with local interactions are much more interesting
than those with long-range interactions.
Also it is meaningful to think of (35) as a one dimensional system whereas for the
system (2) where all the particles interact with each other the concept of
dimensionality is somehow vague. The parameter $ \s$ in (34-36) is used to
treat systems with various boundary conditions. When $ \s = 0 $ we are dealing
with open boundary conditions and when it is non-zero we are dealing with
periodic boundary conditions.
\section{Discussion}
We have shown that the integrable system proposed by Calogero when written in suitable
 coordinates is in fact a system of
non-interacting harmonic oscillators. Our results also show how (in the quantum case)
 the spectrum of this  model can be calculated in a quite simple way. From (24)
it is seen that the energy eigenvalues are given by :
\be E = N\lambda \nu ( n_0 + n_1 + n_2 + . . . .  n_{N-1} ) + N \m \nu n_0 \ee
where $ n_i$ 's are positive integers.
At first sight, it may seem disappointing that the solvable systems of [1,2]
are in fact non-interacting harmonic oscilators in a different guise. However
one should note that after all any {\it integrable} system when written in terms
of the action angle variables will be a {\it non-interacting } system.
Therefore from a rigorous point of view the systems studied in [1,2] deserve to be named
" new solvable systems " although they are too easy to solve when treated properly.
Other integrable systems of the form
$$  H = \sum_{j,k}^{N} {{(p_j p_k)}} \{\lambda + \mu  cos [ \nu ( q_j - q_k)] \} $$
have been proposed by Calogero in [5]. The integrable structure of these systems
have been analysed algebraically in [6,7] , where it has been shown that these systems
are nothing but a system of spins with long range interaction.
\section{Acknowledgements}
I would like to express my gratitude to  F. Ardalan and H. Arfaei for a very
enlightening discussion , to A. Aghamohamadi for drawing my attention to ref. [1]  and
to the participents of IPM seminars for constructive comments.
\newpage

\newpage
{\large \bf References}
\begin{enumerate}

\item  R. Camassa, and D.D. Holm , Phys. Rev. Lett. 7(1993)1661;
R.Camassa,D.D. Holm, and J.M.Hyman, Adv.Appl.Mech. 31(1994)1
\item  F. Calogero : Jour. Math. Phys. {\bf 36} 9 (1995)
\item  F. Calogero and J.F. van Diejen : Phys. Lett. A. {\bf 205} (1995) 143
\item  L. D. Faddeev ; Integrable Models in 1+1 dimensional quantum Fields
theory (Les Houches Lectures 1982), Elsevier, Amsterdam (1984)
\item  F. Calogero : Phys. Lett. A. {\bf 201} (1995) 306-310
\item  V. Karimipour " Algebraic and Geometric Structure of the Integrable Models
recently proposed by Calogero, IPM preprint Feb.96 , Tehran.
\item  V. Karimipour " Relation of The New Calogero Models and xxz Spin Chains "
IPM preprint Feb.96 , Tehran. hep-th/9603039

\end{enumerate}
\end{document}